\documentclass[aps,prl,showpacs,twocolumn,superscriptaddress,longbibliography]{revtex4-1}
\usepackage[pdftex]{graphicx}
\usepackage[pdftex,bookmarks,colorlinks,breaklinks]{hyperref}  
\hypersetup{linkcolor=red,citecolor=blue,filecolor=dullmagenta,urlcolor=blue} 
\newcommand{\var}{\textrm{var}}
\begin{document}
\title{Channel cross-correlations in transport through complex media}
\author{Stefan Gehler}
\affiliation{Fachbereich Physik, Philipps-Universit\"{a}t Marburg, Renthof 5, D-35032 Marburg, Germany}
\author{Bernd K\"{o}ber}
\affiliation{Fachbereich Physik, Philipps-Universit\"{a}t Marburg, Renthof 5, D-35032 Marburg, Germany}
\author{Giuseppe Luca Celardo}
\affiliation{Dipartimento di Matematica e Fisica and Interdisciplinary Laboratories for Advanced Materials Physics, Universit\`a Cattolica, via Musei 41, 25121 Brescia, Italy}
\affiliation{Istituto Nazionale di Fisica Nucleare,  Sezione di Pavia, via Bassi 6, I-27100,  Pavia, Italy}
\author{Ulrich Kuhl}
\thanks{Corresponding author, email: ulrich.kuhl@unice.fr}
\affiliation{Universit\'{e} Nice Sophia Antipolis, CNRS, Laboratoire de Physique de la Mati\`ere Condens\'ee, UMR 7336 Parc Valrose, 06100 Nice, France.}
\affiliation{Fachbereich Physik, Philipps-Universit\"{a}t Marburg, Renthof 5, D-35032 Marburg, Germany}
\date{\today}

\begin{abstract}
Measuring transmission between four antennas in microwave cavities, we investigate directly the channel cross-correlations $C$ of the cross sections $\sigma^{ab}$ from antenna at $\vec{r}_a$ to antenna $\vec{r}_b$. Specifically we look for the $C_\Sigma$ and $C_\Lambda$, where the only difference is that $C_\Lambda$ has none of the four channels in common, whereas $C_\Sigma$ has exactly one channel in common. We find experimentally that these two channel cross-correlations are anti-phased as a function of the channel coupling strength, as predicted by theory. This anti-correlation is essential to give the correct values for the universal conductance fluctuations. To obtain a good agreement between experiment and predictions from random matrix theory the effect of absorption had to be included.
\end{abstract}

\maketitle

Quantum transport in mesoscopic systems is at the center of interest in different fields for its relevance to quantum devices and information technology~\cite{bee97,dni08}.
Quantum transport in mesoscopic systems occurs when the coherence length is larger than the system size~\cite{mel04b}.
From this point of view microwave experiments are an extraordinary laboratory for testing quantum transport in the mesoscopic regime, since the coherence length is basically infinite.
Within this regime the transport is ballistic when the mean free path $l$ is larger than the system size $L$ and it is diffusive when $l<L<\xi$, where $\xi$ is the localization length.
One of the prominent examples of quantum phenomena in mesoscopic systems is given by the universal conductance fluctuations (UCF)~\cite{lee85b,alt85,bee97}.
In the UCF regime, the dimensionless variance of the conductance is independent of sample size or disorder strength and it attains the universal value of $1/8$ in the ballistic regime and of $2/15$ in the diffusive regime.
These universal values are valid in the limit of large number of channels in the external leads and of perfect coupling between the leads and the disordered systems.
On the other side, the transport properties in the mesoscopic regime are strongly affected by the coupling strength with the leads~\cite{sor09,cel08a} and by absorption or dephasing.
The coupling strength with the channels can be tuned in several relevant systems, such as quantum dots or microwave cavities~\cite{bee97}, and the number of channels can
be small.

\begin{figure}
\includegraphics[width=0.45\textwidth]{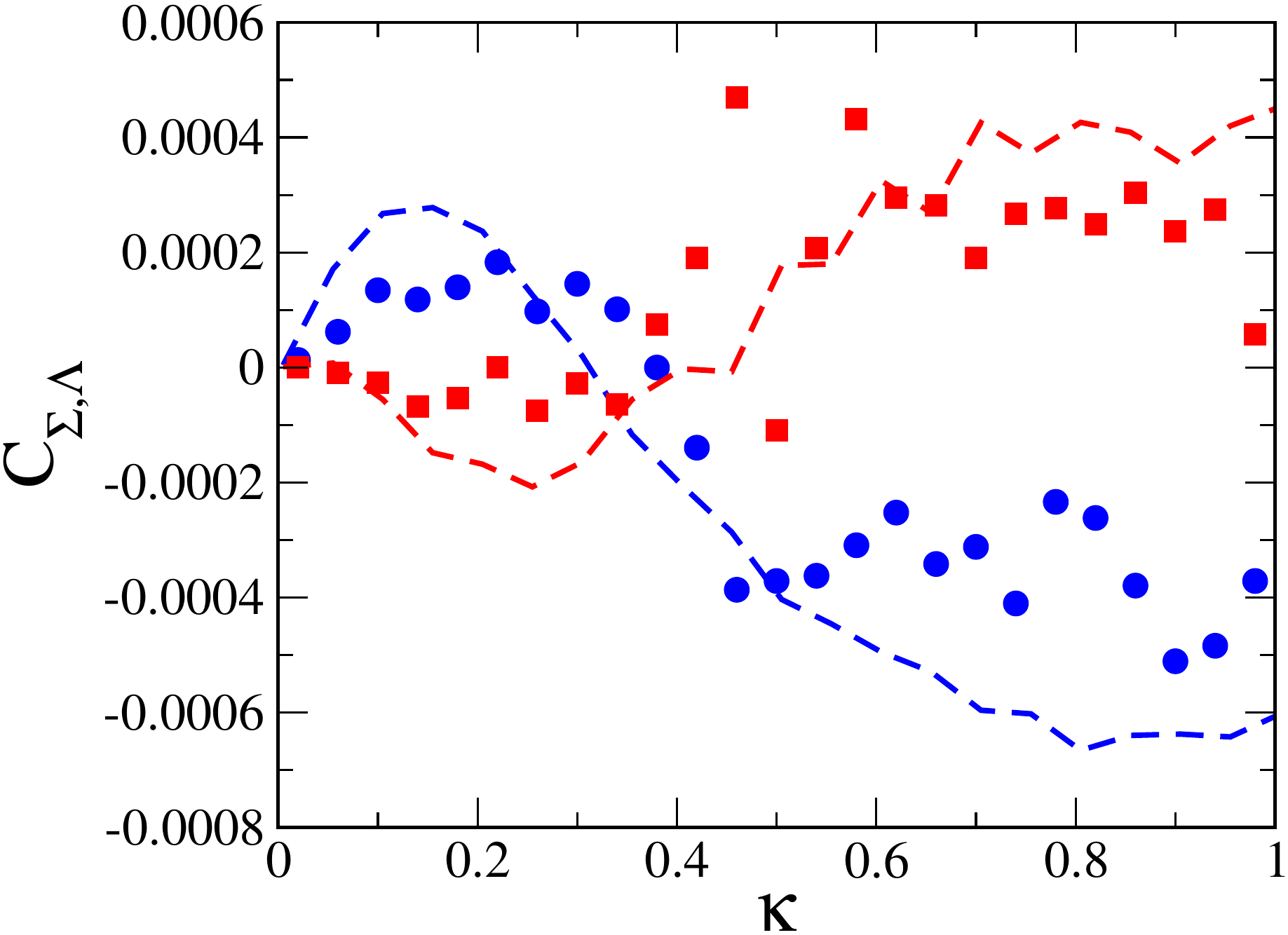}
\caption{\label{fig:Corr}(color online)
Correlation functions $C_{\Sigma}$ (blue dots) and $C_{\Lambda}$ (red squares) as a function of the mean coupling $\kappa$ using a frequency average of 100\,MHz and all measured samples. Symbols are experimental data. Numerical data (dashed curves) have been obtained from a GOE matrix $200\times200$ with $M=4$ channels, using $10^4$ ensemble realizations and $196$ fictitious channels to mimic absorption. In order to have agreement with the experimental results it was essential to take into account the appropriate absorption for each mean coupling $\kappa$, see text.}
\end{figure}

At the core of UCF lies the strong correlations between the transmissions through different channels.
These correlations are the signature of coherent transport, indeed for incoherent transport the correlations are zero.
Channel correlations play an important role in mesoscopic transport~\cite{sor09}, but their direct measurement is a difficult task.
Here we present the results obtained from the analysis of transport through a chaotic microwave cavity attached to four antennas, each playing the role of an independent channel.
The coupling strength $\kappa$ to the antennas are the same and varies with frequency. Channel correlations have been studied as a function of such coupling (see Fig.~\ref{fig:Corr}).
The channel correlations have been extracted from the cross-sections (transmissions, $a\ne b$) from channel $a$ to channel $b$, $\sigma^{ab}$. As we are interested in UCF we concentrate here only on the transmissions.
Channel correlations, as a function of the coupling strength $\kappa$, are defined as:
\begin{equation}\label{eq:main5}
C(\kappa)= \overline{\langle \sigma^{ab}(\kappa)\sigma^{cd}(\kappa)\rangle} - \overline{\langle \sigma^{ab}(\kappa)\rangle \langle \sigma^{cd}(\kappa)\rangle}.
\end{equation}

The $\langle\cdots\rangle$ represents an average over frequency and system configurations and the $\overline{\cdots}$ over different combinations of the indices.
There are two important correlations (\ref{eq:main5}) for the transmission ($a\ne b$ and $c\ne d$).
The so-called $C_{\Sigma}$-correlation has exactly one identical index in the first and second factor.
So either $a=c$ or $b=c$ (or either $a=d$ or $b=d$).
The $C_{\Lambda}$-correlation forbids any equal index.
In Ref.~\cite{sor09}, using Random Matrix Theory (RMT), it was shown that channel correlations strongly depends on the coupling strength between the external environment and the disordered system and on the type of correlations considered ($C_{\Sigma}, C_{\Lambda}$).
In Fig.~(\ref{fig:Corr}) channels correlations obtained experimentally as a function of the coupling strength are shown (symbols) in comparison with numerical results from RMT (dashed curves).
A typical antiphased behavior of the two types of correlations, first predicted in Ref.~\cite{sor09}, is clearly shown.
One of the main challenges to study correlations between different transmission channels was to take into account the role of absorption which inevitably occurs in the experimental set up.
Once the effects of absorption was taken into account, a good agreement was found between theoretical and experimental results.
Usually in electronic transport, also the role of dephasing should be taken into account.
While absorption refers to probability losses, dephasing can occur also in absence of probability losses and it implies that phase coherence is not maintained.
Often dephasing shows similar effects in physical observables as absorption, like broadening of resonances, but it has a different nature, which can be for example also detected in Fano resonances~\cite{baer10a}.
As a result dephasing does not necessarily affect the value of conductance but has a large impact on correlations.
In contrast to electronic transport, in microwave experiment dephasing is practically zero.

\emph{---Theoretical model---}
Statistical properties of a chaotic cavity with time reversal symmetry can be modeled by a GOE Hamiltonian matrix $H$~\cite{men03a,kuh05a,bee97}, where matrix elements are Gaussian random variables, $\langle H_{i,j}^2 \rangle = 1/N$ for $i \ne j$ and $\langle H_{i,j}^2 \rangle = 2/N$ for $i = j$.
According to the well developed formalism~\cite{men03a,kuh05a,mah69,aga75,ver85a,sok88,sok89,sok92,rot91,rot15}, an open system can be described in terms of the effective non-Hermitian Hamiltonian ${\cal H}$,
\begin{equation}\label{eq:Hnon}
{\cal H}= H - \frac{i}{2}\, W\,; \qquad W_{ij}=\sum_{c=1}^M A_{i}^{c}A_j^c,
\end{equation}
where in our case $H$ is a GOE random matrix.
The non-Hermitian part is defined by $W$, standing for the coupling between $N$
intrinsic states $|i\rangle,|j\rangle,$ and $M$ open decay channels labeled as $a,b,c\dots$.
The factorized structure of $W$ is dictated by the unitarity of the scattering matrix.
We restrict ourself to time-invariant systems, therefore, the transition amplitudes $A_{i}^{c}$ between intrinsic states $|i\rangle$ and channels $c$ can be taken as real quantities.
In our study the amplitudes $A_i^c$ are assumed to be random independent Gaussian variables with zero mean and variance~\cite{ver85a,cel07,cel08b,zel04,vol06}
\begin{equation}\label{eq:2}
\langle A_i^c A^{c'}_j\rangle=\delta^{ij}\delta^{cc'}\,\frac{\gamma^{c}}{N}.
\end{equation}
Taking the non-Hermitian matrix (\ref{eq:Hnon}) one can construct the scattering matrix:
\begin{equation}\label{eq:5}
S^{ab}(E)=\delta^{ab}-i{\cal T}^{ab}=\delta^{ab}-i\sum_{i,j}^N A_i^a\left(\frac{1}{E-{\cal H}}
\right)_{ij} A_j^b,
\end{equation}
where ${\cal T}^{ab}$ are the scattering amplitudes determining the cross section $\sigma^{ab}=|{\cal T}^{ab}|^2$ between channel $a$ and channel $b$.

The transmission coefficient $T^c$ through channel $c$ can be written as:
\begin{equation}\label{eq:Tc}
T^c(E)= 1-\left|\left\langle S^{cc} (E)\right\rangle\right|^2= \frac{4 \kappa^c}{(1+\kappa^c)^2}
\end{equation}
where $\kappa^c$ determes the strength of the coupling to the continuum and can be written as
\begin{equation}\label{eq:3a}
\kappa^{c}=\frac{\pi\gamma^{c}}{2ND},
\end{equation}
where $D$ is the mean level spacing.
Here we consider the case of $M$ equiprobable channels, $\kappa^{c}=\kappa$.
We performed all our numerical calculations for the case of GOE ensemble at the center of the energy band, so that $D=\pi/N$, and $\kappa=\gamma/2$.

According to standard scattering theory, the cross section $\sigma^{ab}$ can be computed from the product of the probabilities to enter via channel $a$ and to exit via channel $b$.
The probability to enter through channel $a$ is given by $T^a$, Eq.~(\ref{eq:Tc}) while the probability to exit channel $b$ can be computed by considering all the possible ways to exit the system.
For equivalent channels ($T=T^a$) the probability to exit via any other channel than $a$ is proportional to $(M-1)T$, while the probability to exit from the same channel is proportional to $FT$, where $F$ is the enhancement factor~\cite{cel07,cel08b}.
For the GOE ensemble we have $F=2$.
So that the probability to exit from channel $b$ can be computed as $T^b/(\sum_{c \ne a} T^c +FT^a)$, so that we obtain for equivalent channels~\cite{cel07,cel08b},
\begin{equation}\label{eq:sigma}
\langle \sigma_{{\rm fl}}^{ab} \rangle=\frac{T}{F+M-1};\qquad a \ne b.
\end{equation}

\emph{---Universal conductance fluctuations---}
We can define conductance in analogy with mesoscopic physics, considering $M/2$ $b-$channels as {\it left channels} corresponding to incoming waves, and other $M/2$ $a-$channels, as {\it right channels} for outgoing waves.
Then the conductance $G_0$ can be written as:
\begin{equation}\label{eq:G}
G_0=\sum_{a=1}^{M/2} \sum_{b=M/2 +1}^M \sigma^{ab}\,.
\end{equation}

The properties of the conductance are entirely determined by the inelastic cross-sections, $b\neq a$, and for equivalent channels the average conductance reads (see Eq.~(\ref{eq:sigma})),
\begin{equation}\label{eq:G-av}
\langle G_0 \rangle =\frac{M^2}{4}\langle\sigma^{ab}\rangle=
\frac{M^2}{4}\frac{T}{M-1+F}\,.
\end{equation}

The variance $\var(G_0)$, for perfect coupling, $\kappa=1$, and very large number of channels, $M\gg 1$, equals the famous value $1/8$ for ballistic transport, which is the regime of our experiment.
That the variance is independent of the mean conductance is one of the striking striking effects of universal conductance fluctuations.
The more general expression for finite number of channels (for $\kappa =1$ and the GOE) can be found in Refs.~\cite{mel04b,bee97}:
\begin{equation}\label{eq:vgM}
\var(G_0)=2\frac{(M/2)^2[(M/2)+1]^2}{M(M+3)(M+1)^2}.
\end{equation}
To obtain the correct final expression one has to take the cross section correlations of the kind shown in Eq.~(\ref{eq:main5}) into account appropriately~\cite{lee87,fen88,fre88,mel88}.
They enter into the variance in the following form~\cite{sor09}:
\begin{equation}\label{eq:var+corr}
\var(G_0) = \frac{M^2}{4}\left(\frac{T}{F+M-1}\right)^2 + N_{\Sigma} C_{\Sigma} + N_{\Lambda} C_{\Lambda},
\end{equation}
where $N_{\Sigma}=L(M-2), \,N_{\Lambda}=L(L-M+1),\, L=M^2/4$ and $C_{\Sigma},C_{\Lambda}$ are the two types of correlations discussed in the introduction.
As one can see, if one neglects the correlations, the first term gives $1/4$ (for $T=1$ and $M\gg 1$), instead of the value $1/8$.
This result clearly demonstrates the crucial role of correlations determining the conductance fluctuations (see also~\cite{gar02}).

\emph{---Absorption---}
\begin{figure}
\includegraphics[width=0.45\textwidth]{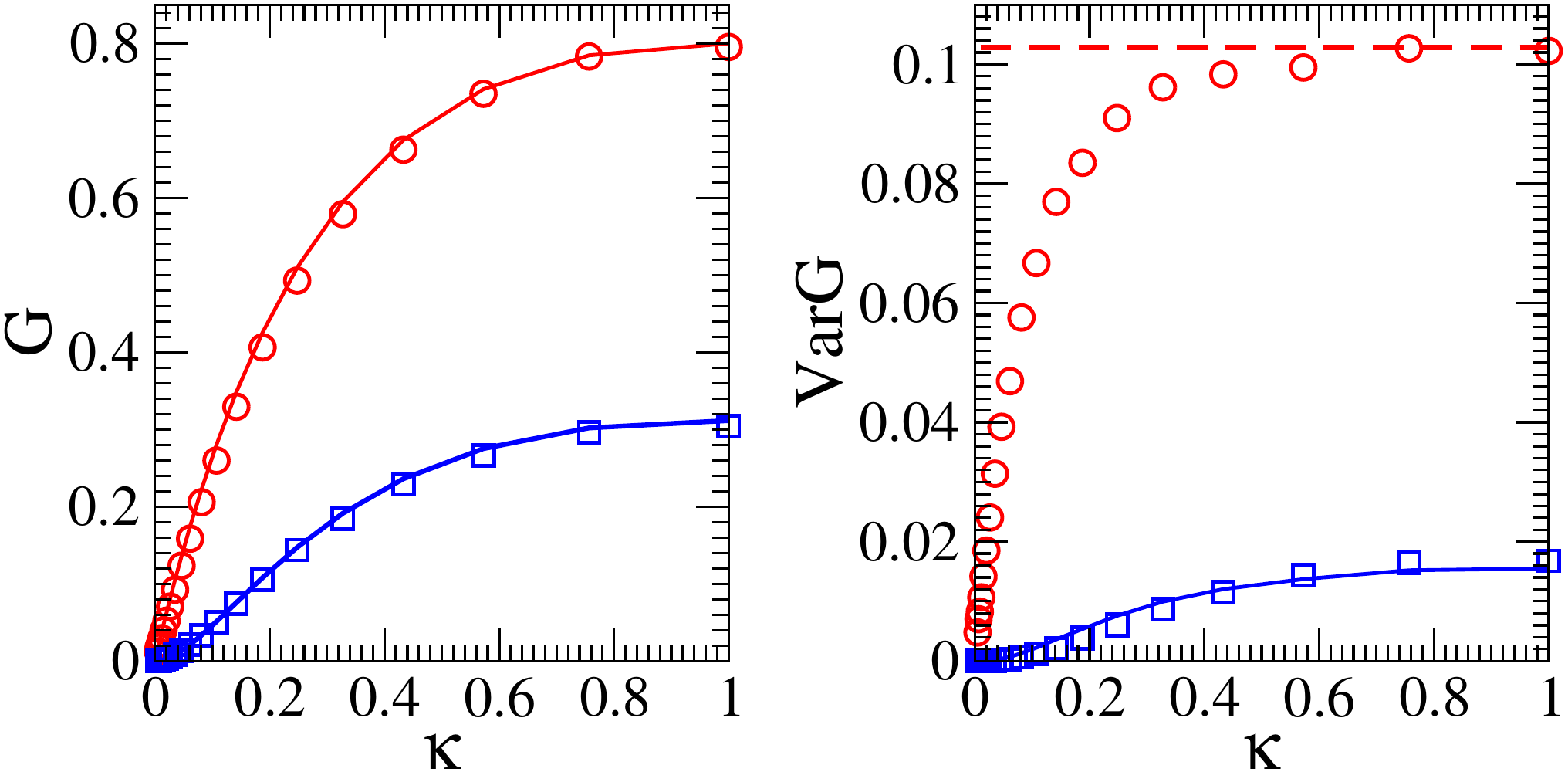}
\caption{\label{fig:ABSth}(color online)
Conductance (left) and its fluctuations (right) as function of coupling $\kappa$.
Red circles refer to values without absorption, while blue squares refer to the values with constant absorption $\kappa_A=0.02$.
Left panel: analytical result for the conductance (curves), see Eq.~(\ref{eq:Gth}), are shown together with numerical results (symbols).
Right panel: numerical results for $\var(G)$ without absorption (red circles) and with absorption (blue squares) are shown.
The full blue curve has been obtained by multiplying the data of $var(G)$ without absorption by the absorption factor $q_a^2$, see Eq.~(\ref{eq:qa}).
The dashed horizontal lines refer to theoretical value of the variance of $G$ at perfect coupling, see Eq.~(\ref{eq:vgM}).
Numerical data have been obtained using a GOE matrix with $N=200$ elements, $M=4$ transmitting channels and $N-M$ absorbing channels.
$10^{4}$ ensemble realization have been used to obtain the numerical data.
}
\end{figure}
In experiments absorption plays an important role and can included in the effective non-Hermitian Hamiltonian by adding many weakly coupled decaying channels.
Therefore we introduce in addition to the $M$ transmitting channels $N_A$ weakly coupled fictitious channels to describe absorption. We assume $N_A$ is large and all fictitious channels have equal tiny coupling $\kappa_A$.
Thus transmission through these channels can be computed from Eq.~(\ref{eq:Tc}) and we have:
\begin{equation}\label{eq:TW}
T_W= 4 N_A \kappa_A\,.
\end{equation}

We can estimate the effect of absorption on the conductance and its variance.
Assuming equal coupling to the transmitting channels, for an incident wave in channel $a$, the transmitted part is $T$.
The part which will escape through the other transmitting channels is proportional to $(M-1)T$, the part which gets lost in the absorbing channels is proportional to $T_W$ and the part which escapes back from the same entering channels is proportional to $FT$, where $F$ is the enhancement factor.
Following the same reasoning done for Eq.~(\ref{eq:sigma}), we can now evaluate the cross section between channels $a$ and $b$ in presence of absorption as:
$\langle \sigma^{ab}\rangle=\frac{T^2}{(M-1+F)T+T_W}$, which can be written as:
\begin{equation}\label{eq:qa}
\langle \sigma^{ab}\rangle = \frac{T}{M-1+F} \left(1-\frac{T_W}{(M-1+F)T+T_W} \right) = q_a \langle \sigma^{ab}_0\rangle\,,
\end{equation}
where $\sigma^{ab}_0$ is the cross section in absence of absorption, see Eq.~(\ref{eq:sigma}), and the absorbing factor is given by
\begin{equation}\label{eq:qadef}
q_a=\frac{\langle \sigma^{ab}\rangle}{\langle \sigma_0^{ab}\rangle} =1-\frac{T_W}{(M-1+F)T+T_W}
\end{equation}
For the conductance in presence of absorption we have
\begin{equation}\label{eq:Gth}
\langle G \rangle = q_a \langle G_0 \rangle
\end{equation}
while for its variance $\var(G)= q^2_a \var(G_0)$.
In Fig.~(\ref{fig:ABSth}) we compare numerical results obtained with random matrix approach with the theoretical estimates of the effect of constant absorption.
Both for the case of conductance and of its variance, the absorbing factor defined in Eq.~(\ref{eq:qa}) accounts nicely for the effects of absorption.

\emph{---Experimental set-up---}
\begin{figure}
\includegraphics[width=3.8cm]{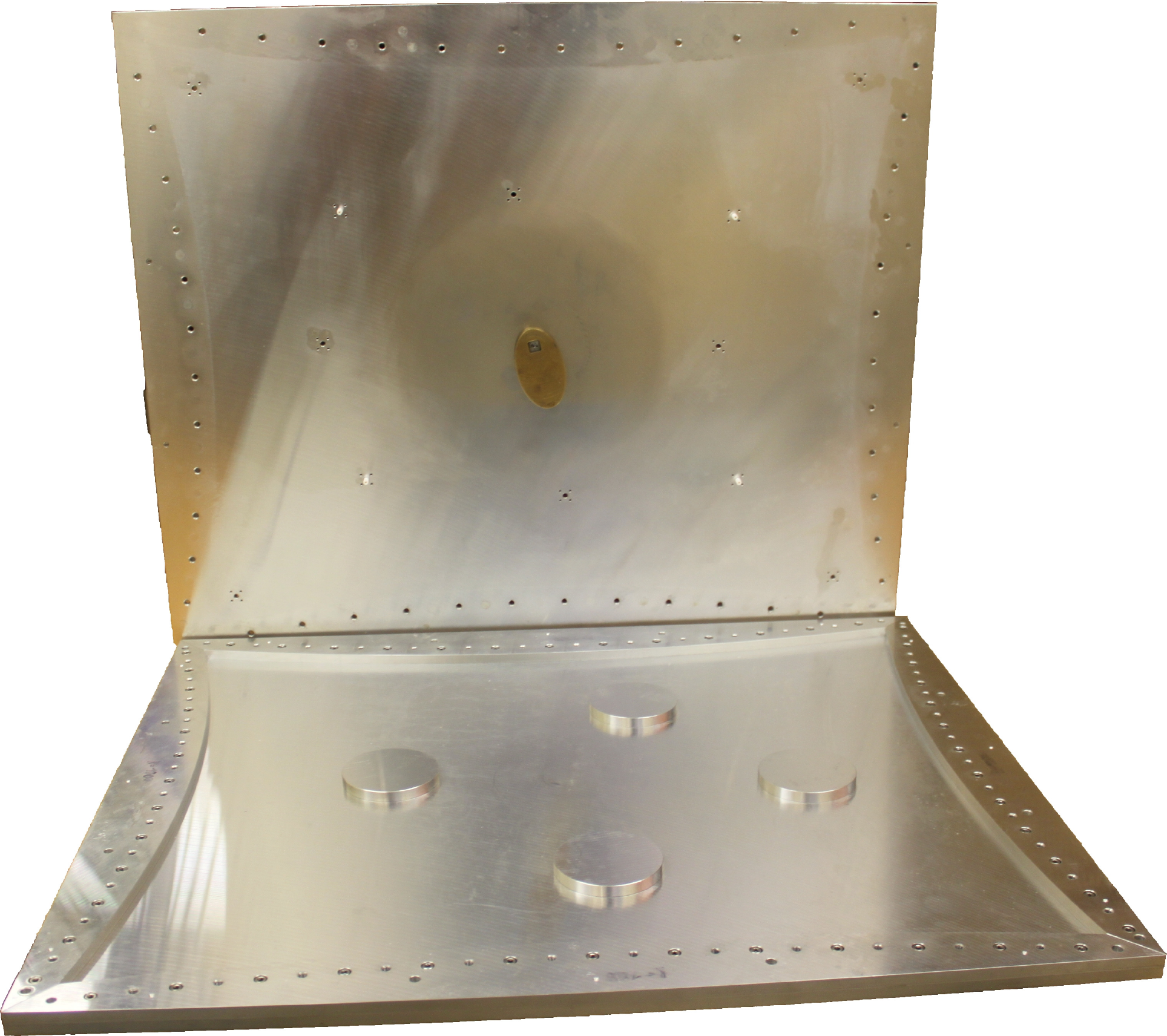}
\raisebox{3.cm}[0pt][0pt]{\hspace{-4cm}(a)}\hspace{3.5cm}
\raisebox{3.cm}[0pt][0pt]{(b)}\hspace{-0.2cm}
\includegraphics[width=4.2cm]{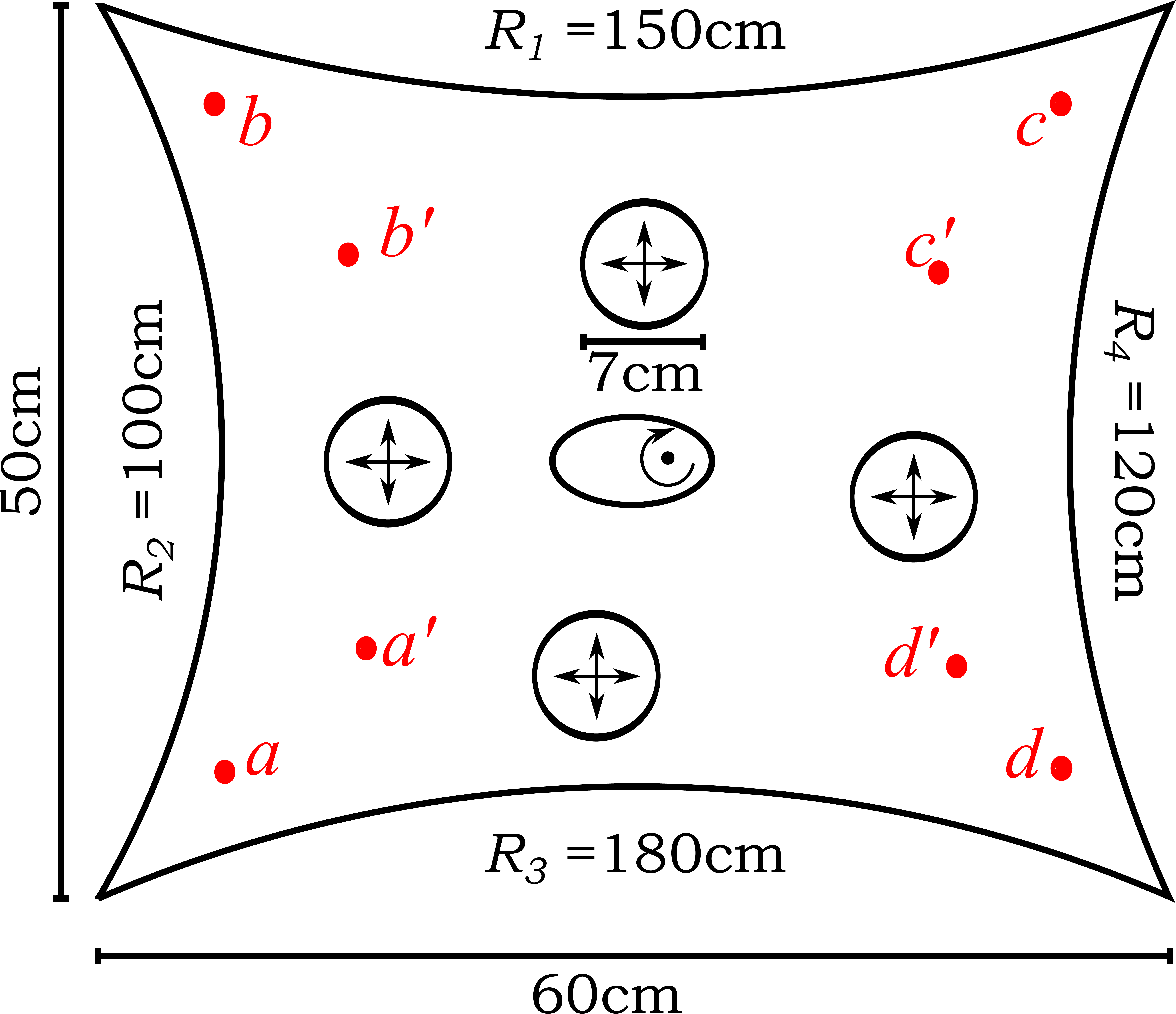}
\caption{\label{fig:ExpSetupCrossCorr}(color online)
(a) A photograph of the setup and (b) sketch of the microwave resonator with a height of 8\,mm.}
\end{figure}
Fig.~\ref{fig:ExpSetupCrossCorr}(a) shows a photograph of the used setup with a raised top plate.
The billiard is made out of aluminium and in (b) one can see the shape of the billiard, which generates a classical chaotic behavior.
The four circle insets with variable positions were introduced to avoid direct processes between the four antennas $a$, $b$, $c$, and $d$.
For the ensemble average, different positions of the circles and rotations of the ellipse were chosen (in the photograph the brass ellipse can be seen at the center of the top plate).
For each configuration of the circles we measured the scattering matrix for $200$ different angles (in steps of $1.8^\circ$).

\begin{figure}
\includegraphics[width=\linewidth]{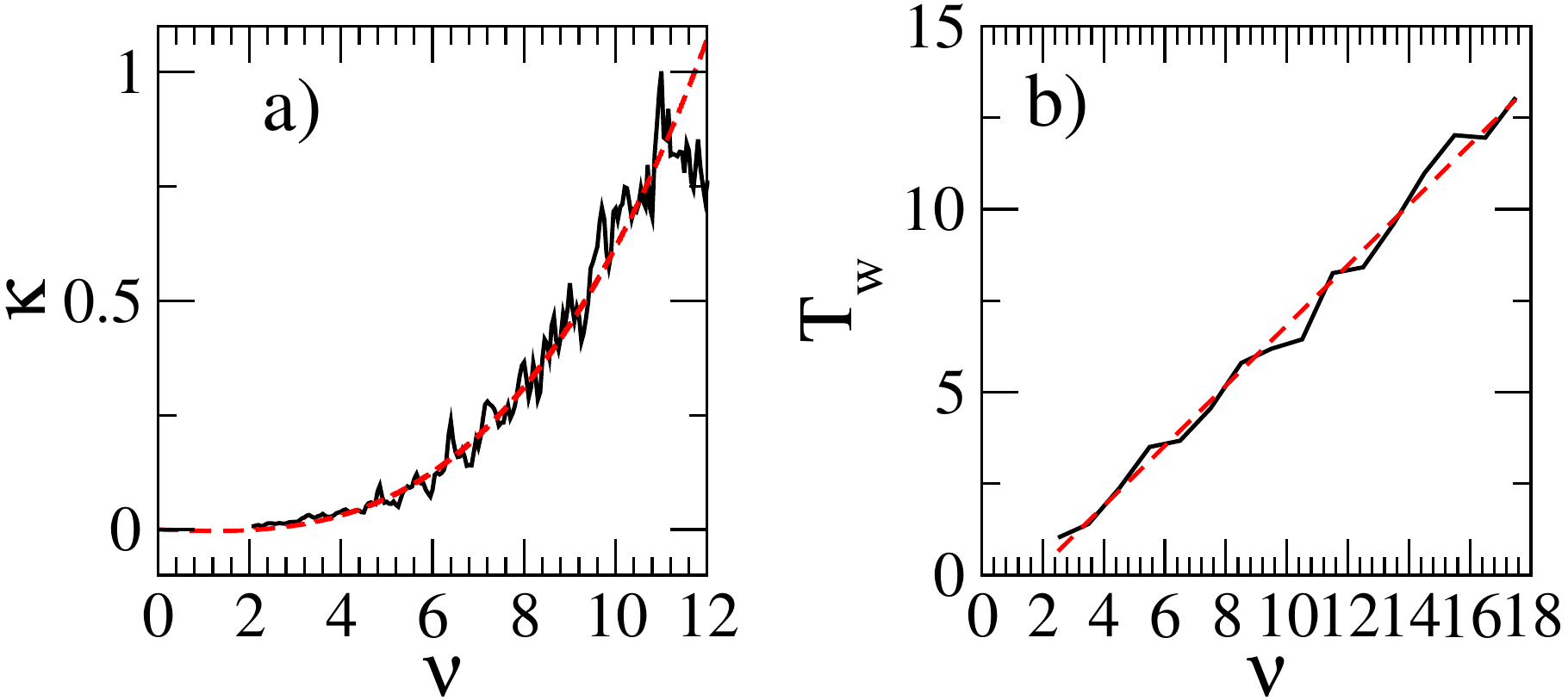}
\caption{\label{fig:autocorr}(color online)
(a) Coupling $\kappa$ as a function of the frequency $\nu$.
The dashed red curve show a cubic approximation ($\kappa=-0.0043 \nu +0.0006 \nu^2 +0.0006 \nu^3$).
(b) Wall absorption ($T_W$ is the transmission to the wall channels) as a function of the frequency. The dashed red curve shows a linear approximation ($T_W=-1.4 +0.82 \nu$).}
\end{figure}

For these measurements a four-port vector network analyzer (Agilent E5071C) was used, which allows to measure all components of the ($4\times 4$) $S$-matrix without the need of a permanent change of cables between the different antennas.
The $S$-matrix was determined in a frequency range from 2 to 18\,GHz in steps of 0.2\,MHz.
The height of the billiard is $h=8$\,mm and so for frequencies $\nu < \frac{c}{2\cdot h} \approx 18.7$\,GHz the describing Helmholtz equation is fully equivalent to the two dimensional Schr\"{o}dinger equation~\cite{stoe90}.
The four attached antennas extends 5\,mm into the resonator and were chosen to be as similar as possible.
Transmissions between different channels have been computed as a function of the frequency of the incoming wave $\sigma^{ab}(\nu)=|S^{ab}(\nu)|^2, \textnormal{ with }a\neq b$.
We are interested in the cross section as a function of the coupling $\kappa$ and not as a function of the frequency.
In our experiments the mean scattering amplitude $\langle S^{ii}(\nu)\rangle$ is not always real due to global phase shifts induced mainly by the antennas~\cite{kuh07b}.
Following the procedure used in~\cite{koeb10}, we use the complex version of Eq.~(\ref{eq:Tc}), obeying equation
\begin{equation}\label{eq:TiComplex}
T^i=1-|\langle S^{ii} \rangle|^2=4\textnormal{Re}(\kappa^i)/|1+\kappa^i|^2,
\end{equation}
thus minimizing the effect coming from the phase of $\langle S^{ii} \rangle$.
By averaging over windows of 100\,MHz, all samples, and the 4 antennas, the coupling is extracted according to Eq.~(\ref{eq:TiComplex}).
The results are plotted in Fig.~(\ref{fig:autocorr}) left panel.
With this averaged coupling dependence we have a mapping between the frequency axis and the coupling axis.

Figure~\ref{fig:autocorr} right panel, shows the dependence of wall absorption ($T_W$) of the used aluminium billiard extracted from the exponential decay of the auto-correlation functions~\cite{sch03a}.

Following Ref.~\cite{sch03a} we can write $T_W=\sum_{a=1}^{N_A} T^a$, where $N_A$ is the number of absorbing channels and $T^a$ is the transmission coefficient for each absorbing channels.
The dependence of transmission coefficients on the coupling is given in Eq.~(\ref{eq:Tc}).
Since the coupling to the absorption channels is small, and assuming uniform coupling, we have: $T_W \approx 4 N_A \kappa_A$.
From Fig.~(\ref{fig:autocorr}) we can obtain by a fitting the dependence of the wall absorption on frequency $\nu$.
Combining this result with the dependence of $\kappa$ on the frequency, we can extract the dependence of $T_W$ on $\kappa$, thus fixing all parameter in our model.

\emph{---Experimental results---}
We measured the 16 elements of the scattering matrix $S^{ab}$ and calculated the cross sections $\sigma^{ab}$.
$\sigma$ as a function of $\kappa$ is averaged first on a grid of step size $\Delta \kappa=1/25$ (see Eq.~\ref{eq:main5}).
Afterwards we averaged over the 200 rotation positions (steps of $1.8^\circ$) and three different positions of the circle.
Now we calculated the correlation function and thereafter averaged over all possible antenna combinations.

\begin{figure}
\includegraphics[width=0.45\textwidth]{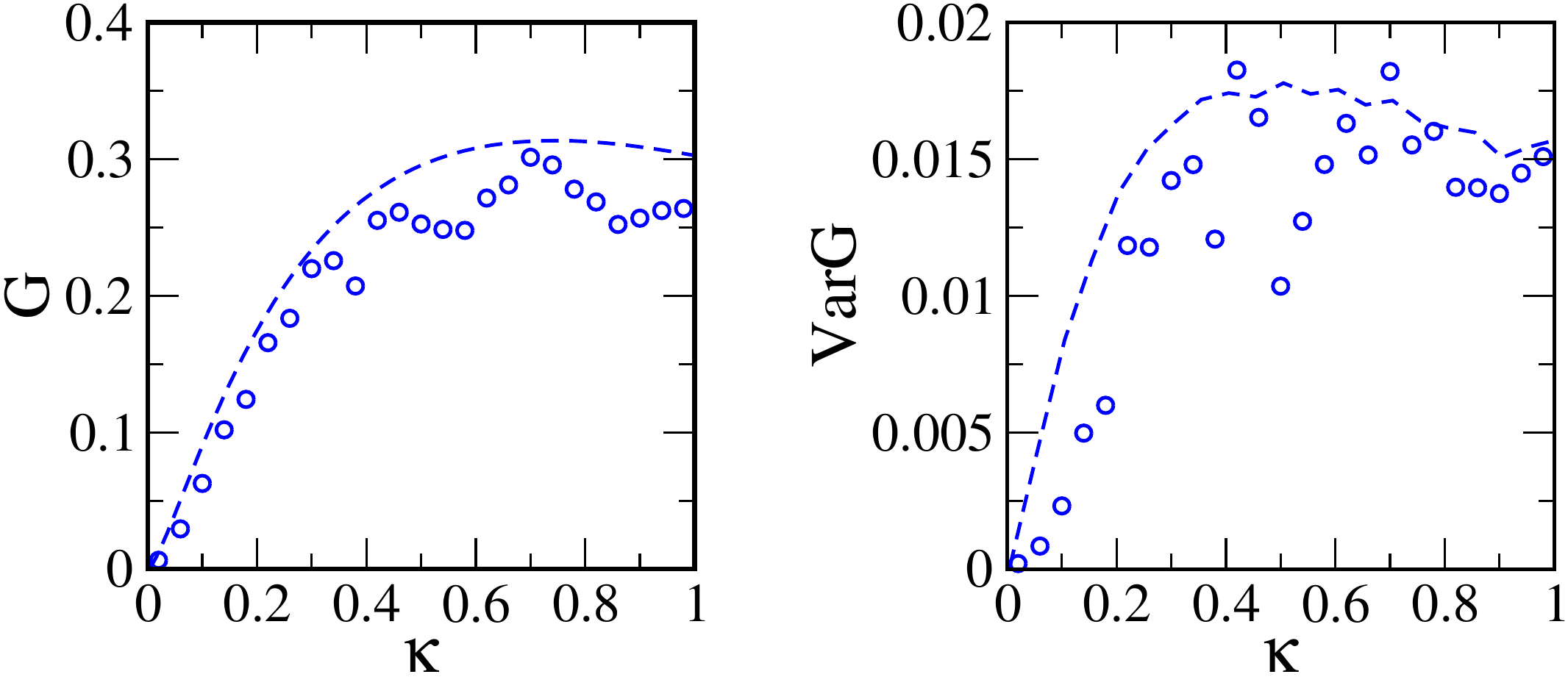}
\caption{\label{fig:CondFluctExp}(color online)
Experimental conductance, (a) and its variance (b) as a function of the mean coupling $\kappa$ for a frequency average of 100\,MHz. The symbols correspond to the experimental data. In (a) the dashed curved is given by the analytical expression Eq.~(\ref{eq:Gth}), whereas in (b) the dashed curve has been obtained by multiplying the numerical results for the case of no absorption by the absorbing factor $q_a$, see Eqs.~(\ref{eq:qa}) and ~(\ref{eq:qadef}).
Numerical data have been obtained from a GOE matrix $200\times200$ with $M=4$ channels, using $10^4$ ensemble realizations.
}
\end{figure}

The experimental results and numerical simulations of the conductance and its variance are shown in Fig.~\ref{fig:CondFluctExp}.
Once the effect of absorption is taken into account a good agreement is obtained between experimental data (symbols) and RMT results (dashed curves), especially as there is no free parameter.

In Fig.~\ref{fig:Corr} the dependence of the two correlations $C_{\Sigma}$ (blue circles) and $C_{\Lambda}$ (red squares) on the channel coupling $\kappa$ are shown.
The dashed lines are results from random matrix numerics obtained by multiplying the channel correlations obtained for zero absorption by the absorption factor squared [see Eq.~(\ref{eq:qa})].
One can clearly see an anti-phased behavior of the two correlation functions.
A slightly better agreement between experiment and numerics is observed when we directly take the values from Fig.~\ref{fig:autocorr} and not the fitted values, also leading to additional fluctuations in the numerical curves.

\emph{---Conclusion---}
Here we report the experimental confirmation of the specific dependence of the channel correlations in the ballistic regime as a function of the coupling strength with the continuum of states in the external environment.
The experimental results reported here shed light on the nature of correlations between different channels which plays a fundamental role in the quantum features of mesoscopic transport.
We showed that the channel correlations have a typical anti-phased coupling dependence.
Such channel correlations constitute a specific signature of coherent transport and their role is essential in understanding universal conductance fluctuations.
Note that channel correlations are relevant also in nuclear physics in the context of Ericson fluctuations, see discussion in Ref.~\cite{eri60,eri63,wei90}.

In order to compare experimental results with theoretical predictions from RMT, the absorption needs to be taken into account.
Using an absorbing factor, which has been computed analytically, a good agreement is found for the averaged conductance and its variance. Apart from the antiphased-structure of the two correlation functions even a quantitative agreement is found between numerical RMT estimates with the experiments, where the absorption has been fixed in advance, the only free parameter in the RMT model.

\begin{acknowledgments}
We thank F.~Izrailev for suggesting this experiment and the Deutsche Forschungsgemeinschaft for the support via the FOR 760.
\end{acknowledgments}

\end{document}